\documentclass[%
twocolumn,
superscriptaddress,
 amsmath,amssymb,
 aps,
prl,
]{revtex4}

\usepackage{graphicx}
\usepackage{dcolumn}
\usepackage{bm}
\usepackage[colorlinks,
           urlcolor=blue,
           linkcolor=blue,
           anchorcolor=blue,
           citecolor=blue
           ]{hyperref}
\usepackage{float}
\usepackage{xcolor}

\begin{document}

\preprint{APS/123-QED}

\title{Raman Linewidth Contributions from Four-Phonon \\ and Electron-Phonon Interactions in Graphene}



\author{Zherui Han}
 \altaffiliation{These authors contributed equally to this work.}
 \affiliation{School of Mechanical Engineering and the Birck Nanotechnology Center,\\
Purdue University, West Lafayette, Indiana 47907-2088, USA.}
\author{Xiaolong Yang}%
 \altaffiliation{These authors contributed equally to this work.}
 \affiliation{Institute for Advanced Study, Shenzhen University, Shenzhen 518060, China.}
\author{Sean E. Sullivan}
 \affiliation{Department of Mechanical Engineering, The University of Texas at Austin, Austin, Texas 78712, USA.}
\author{Tianli Feng}
 \affiliation{Department of Mechanical Engineering, The University of Utah, Salt Lake City, Utah 84112, USA.}
\author{Li Shi}
 \affiliation{Department of Mechanical Engineering, The University of Texas at Austin, Austin, Texas 78712, USA.}
\author{Wu Li}
 \email{wu.li.phys2011@gmail.com}
 \affiliation{Institute for Advanced Study, Shenzhen University, Shenzhen 518060, China.}
\author{Xiulin Ruan}%
 \email{ruan@purdue.edu}
 \affiliation{School of Mechanical Engineering and the Birck Nanotechnology Center,\\
Purdue University, West Lafayette, Indiana 47907-2088, USA.}

\date{\today}

\begin{abstract}
The Raman peak position and linewidth provide insight into phonon anharmonicity and electron-phonon interactions (EPI) in materials. For monolayer graphene, prior first-principles calculations have yielded decreasing linewidth with increasing temperature, which is opposite to measurement results. Here, we explicitly consider four-phonon anharmonicity, phonon renormalization, and electron-phonon coupling, and find all to be important to successfully explain both the $G$ peak frequency shift and linewidths in our suspended graphene sample at a wide temperature range. Four-phonon scattering contributes a prominent linewidth that increases with temperature, while temperature dependence from EPI is found to be reversed above a doping threshold ($\hbar\omega_G/2$, with $\omega_G$ being the frequency of the $G$ phonon).

\end{abstract}

\maketitle



Graphene has been studied~\cite{grapheneExfoliation,grapheneRising2007,grapheneReview2019} as an emerging atomically thin electronic and optoelectronic material and for thermal management~\cite{nature2010photonics,apl2008thermal}. Weak interactions between some acoustic phonon polarizations especially the flexural modes with the electronic and optical phonon excitations can give rise to hot electrons and overpopulated optical phonons~\cite{nl2009graphene,nl2009hotphonons} and limit the heat spreading contribution from low frequency phonons in graphene electronic and optoelectronic devices~\cite{nl2017nonequilibrium}. Meanwhile, the increased population of hot charge carriers can enhance the responsivity of graphene-based photodetectors~\cite{nl2013responsivity}. A detailed understanding of electron-phonon and phonon-phonon interaction is essential to understanding the transport properties and device performance of graphene and other 2D systems. 

Raman spectroscopy provides an useful probe of the electron-phonon and phonon-phonon interactions in solid-state materials such as graphene~\cite{raman1984,beechem2007invited,raman2013}, and it has important implications for phonon anharmonicity. The Raman peak shift and linewidth depend on the coupling of the  Raman-active optical phonon mode with electrons and other phonon polarizations. Prior first-principles studies~\cite{prl2007anharm} have investigated the contributions from electron-phonon and phonon-phonon scatterings to the linewidths of $G$ peak caused by Raman scattering of a zone-center optical phonon in graphene. The intrinsic phonon linewidth $\gamma^{\rm in}$ is expressed as $\gamma^{\rm in}=\gamma^{\rm e-ph}+\gamma^{\rm ph-ph}$ with $\gamma^{\rm e-ph}$ and $\gamma^{\rm ph-ph}$ representing contributions from the electron-phonon (e-ph) and phonon-phonon interactions (ph-ph), respectively. It was predicted that $\gamma^{\rm e-ph}$ decreases while $\gamma^{\rm ph-ph}$ increases with temperature ($T$), and the descending trend of $\gamma^{\rm e-ph}$ would dominate up to 800~K~\cite{prl2007anharm}. Opposite to this theoretical prediction, prior experiments show a monotonically increased linewidths with $T$ in graphite, few-layer graphene, and supported monolayer graphene~\cite{prb2010lifetimeExp,prl2010opExp,prb2011measurement,carbon2013measurement}. This contradiction between theory and experiment underscores the need for an in-depth examination of the relative strength of intrinsic electron-phonon and phonon-phonon interactions in graphene. 

In this study, we employ first-principles methods that explicitly consider higher-order phonon anharmonicity based on recent advances~\cite{4ph2016,4ph2017}. Specifically, we account for both three-phonon scattering contribution ($\gamma^{\rm 3ph}$) and four-phonon scattering contribution ($\gamma^{\rm 4ph}$) in the calculation of $\gamma^{\rm ph-ph}=\gamma^{\rm 3ph}+\gamma^{\rm 4ph}$ without involving fitting parameters that were used in several prior studies~\cite{prb2011measurement,JP2012Temp,BN2016Temp}. We further utilize a recently developed temperature dependent effective potential (TDEP) formalism~\cite{TDEP2013}, which can be combined with four-phonon scattering for a unified treatment~\cite{ravichandran2018unified,xia2018revisiting}, to capture the phonon renormalization effect in graphene. While a prior work~\cite{4phlinewidth2020} suggested that the four-phonon scattering channel is generally important and even dominant in the zone-center optical phonon linewidth in 3D dielectric crystals, our results show that considering the effect of temperature is necessary for accurately predicting $\gamma^{\rm 4ph}$ in pristine graphene and can allow accurate prediction of the Raman peak shift at finite temperatures. In particular, $\gamma^{\rm 4ph}$ in graphene would be greatly overestimated if the effect of temperature on the phonon self-energy is neglected. The calculated linewidth and peak shift agree well with previous experiments of supported graphene and our own measurements of clean suspended monolayer graphene. By considering not only EPI as in previous work~\cite{hasdeo2016fermi} but also the $T$ dependence of EPI in addition to full anharmonicity, our calculations predict that $\gamma^{\rm e-ph}$ changes nonmonotonically with increasing doping level and temperature. \\


All first-principles calculations are implemented in VASP package~\cite{VASP1993} and QUANTUM-ESPRESSO package~\cite{QE}. The ShengBTE code incorporating four-phonon scattering~\cite{shengbte,han2021fourphonon} is then used to calculate ph-ph scattering rates. The EPW package~\cite{epw} is used to calculate e-ph scattering rates. To consider the phonon renormalization effect, we use TDEP to compute $T$-dependent effective interatomic force constants (IFCs). Further computational details are presented in the Supplemental Material~\cite{supply}.

\begin{figure}[h]
    \centering
    \includegraphics[width=3.2in]{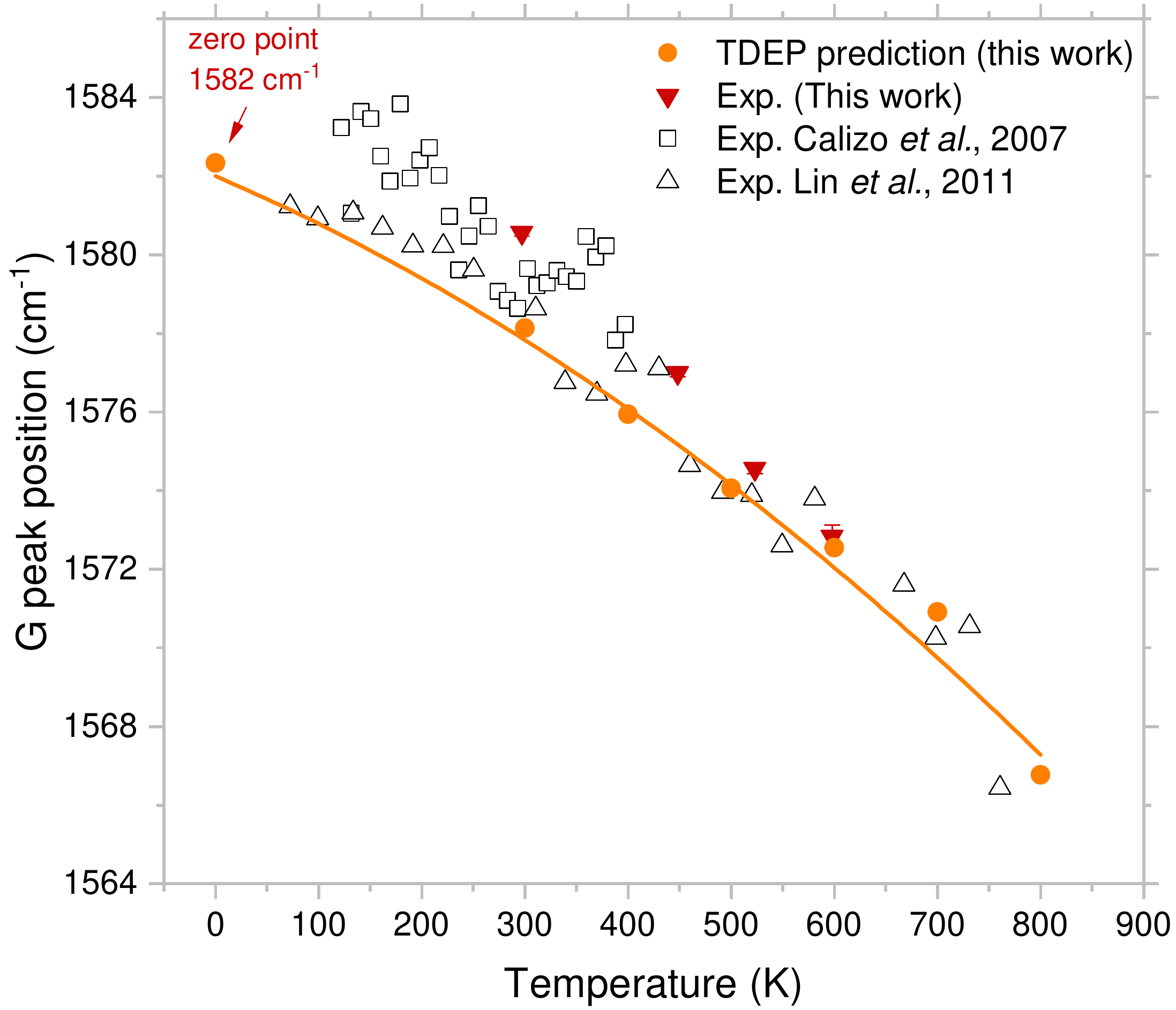}
    \caption{$G$ peak frequency shift of graphene from 0 K to 800 K. Orange solid line is a quadratic fitting to our calculated results at different temperatures using TDEP method. Experiment data are from Ref.~\cite{nanoletter2007ramanfrequency,prb2011measurement} and this work.}
    \label{freqshift}
\end{figure}

We first present our results of phonon anharmonicity in pristine graphene. Figure~\ref{freqshift} presents the first principles predicted Raman $G$ peak frequency, which is equivalent to the frequency of the zone-center $E_{2g}$ mode. The results show good agreement with the available experiments~\cite{nanoletter2007ramanfrequency,prb2011measurement} and our own measurement of monolayer graphene sample grown by chemical vapor deposition and suspended over a circular hole~\cite{nl2017nonequilibrium}, which validates our choice of using the TDEP method to capture the temperature effects in graphene. This $T$ dependence of the Raman $G$ peak is a signature of anharmonicity that comes from both phonon-phonon interactions and lattice thermal expansion. The TDEP method intrinsically includes the impact of  higher-order phonon-phonon interactions on the phonon frequency~\cite{TDEP2018PNAS}. For the thermal expansion contribution, we directly use the first-principles-predicted lattice constants from Ref.~\cite{Marzari2005lattice} in our calculations. Consistent with Ref.~\cite{prl2007anharm}, and despite the negative thermal expansion of graphene, the overall frequency shift still decreases with increasing $T$.

\begin{figure}
    \centering
    \includegraphics[width=3.4in]{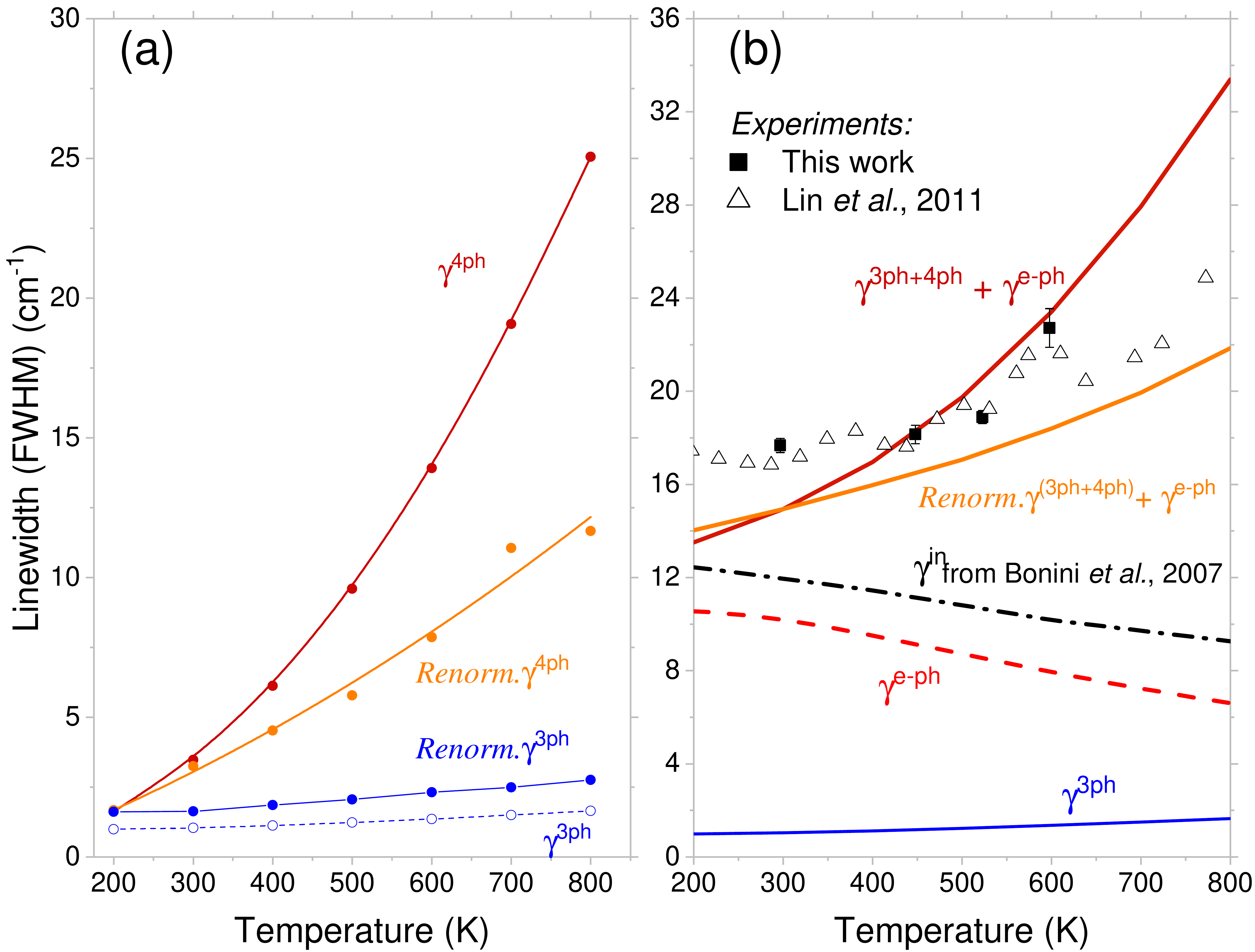}
    \caption{(a) Intrinsic $E_{2g}$ mode linewidth from ph-ph interactions $\gamma^{\rm ph-ph}$, with and without phonon renormalization. Here, the data with '\textit{Renorm.}' at the beginning of the label are those calculated with TDEP to account for the temperature effect, and Renorm.$\gamma^{\rm 4ph}$ is fitted to a quatratic function. (b) Intrinsic $E_{2g}$ mode linewidth of graphene from e-ph and ph-ph contributions. $\gamma^{\rm e-ph}$ is calculated for pristine graphene. Experiments are from Ref.~\cite{prb2011measurement} and our own measurements. A previous prediction~\cite{prl2007anharm} is plotted as the dash-dot black line, and shows an opposite dependence on $T$ compared to measurements.}
    \label{gamma}
\end{figure}

We next shift our focus to the calculation of phonon linewidth, which is related to the phonon-phonon scattering rate $\tau^{-1}$ as $\gamma^{\rm ph-ph}=\frac{\tau^{-1}}{2\pi}$. 
Figure~\ref{gamma}(a) presents our calculated $\gamma^{\rm 3ph}$ and $\gamma^{\rm 4ph}$, which are  expressed in the full width at half maximum (FWHM). The results convey two important insights. First,  $\gamma^{\rm 4ph}$ is dominant over $\gamma^{\rm 3ph}$, even at room temperature. With rising $T$, $\gamma^{\rm 3ph}$ only slightly increases while $\gamma^{\rm 4ph}$ grows dramatically. Based on this finding, neglection of four-phonon scattering is the main cause of the opposite $T$ dependence calculated in Ref.~\cite{prl2007anharm} compared to experiments, as given in Fig.~\ref{gamma}(b) (dash dot black curve). Second, we note that the modification of the phonon self-energy with $T$ is necessary for accurate calculation of $\gamma^{\rm 4ph}$, which exhibits a different $T$ dependence compared to $\gamma^{\rm 3ph}$. Without considering the phonon renormalization effect, the calculation would overestimate the four-phonon scattering rates especially at higher temperatures, as shown by the comparison between the red and orange lines in Fig.~\ref{gamma}(a). In comparison, $\gamma^{\rm 3ph}$ is relatively insensitive to temperatures (see comparison between the dashed and solid blue lines in Fig.~\ref{gamma}(a)). Similarly, a recent work on graphene~\cite{Gu2019temperature} based on an optimized Tersoff potential suggests that fourth-order IFCs show much stronger dependence on $T$ than third-order IFCs.


For pristine graphene, the EPI contribution to FWHM is given by Fermi’s golden rule~\cite{prl2007anharm}:

\begin{equation}
    \begin{array}{r}
    \gamma^{\mathrm{e}-\mathrm{ph}}(T)=\frac{4 \pi}{N_{\mathbf{k}}} \sum\limits_{\mathbf{k}, i, j}\left|g_{(\mathbf{k}+\mathbf{q}) j, \mathbf{k} i}\right|^{2}\left[f_{\mathbf{k} i}(T)-f_{(\mathbf{k}+\mathbf{q}) j}(T)\right] \\
    \times \delta\left[\epsilon_{\mathbf{k} i}-\epsilon_{(\mathbf{k}+\mathbf{q}) j}+\hbar \omega_{\mathbf{q}}\right]
    \end{array}
    \label{eq.e-p}
\end{equation}
where $\omega_{\mathbf{q}}$ is the phonon frequency, the sum is on $N_{\mathbf{k}}$ electron vectors $\mathbf{k}$. $g_{(\mathbf{k}+\mathbf{q}) j, \mathbf{k} i}$ is the e-ph coupling matrix element for a phonon with wave vector $\mathbf{q}$ exciting an electronic state $|\mathbf{k} i\rangle$ with wavevector $\mathbf{k}$ into the state $|(\mathbf{k}+\mathbf{q}) j\rangle$. $f_{\mathbf{k} i}(T)$ is the Fermi-Dirac occupation for an electron with energy $\epsilon_{\mathbf{k} i}$, $\delta$ is the Dirac delta used to describe the energy selection rule.

With the above results, Figure~\ref{gamma}(b) shows our calculated intrinsic phonon linewidth $\gamma^{\rm in}$ of pristine graphene. The red solid line in Fig.~\ref{gamma}(b) shows that $\gamma^{\rm in}$ calculated without temperature corrections would be well above the experimental data. With four-phonon scattering and phonon renormalization accounted for, the obtained renormalized linewidth agrees reasonably well with prior experiments and our own measurements of suspended monolayer graphene. While the absolute FWHM values of our calculation results fall slightly below the experimental data for pristine graphene, this discrepancy could be explained by the finite instrument resolution of the spectrometers used in the experiments, which is on the order of a few cm$^{-1}$ in our measurements and would broaden the measured linewidth. In contrast to the dash dot black line calculated in previous studies~\cite{prl2007anharm}, the phonon linewidth for the $E_{2g}$ mode is not completely dominated by $\gamma^{\rm e-ph}$. Rather, its descending trend would be compensated and then outweighed by the growing $\gamma^{\rm ph-ph}$, which is mainly due to the increasing four-phonon scattering rates at higher temperatures. Thus, at all temperatures $\gamma^{\rm in}$ exhibits an increasing trend. Our calculations successfully explain this $T$ dependence observed in experiments.

Our calculations above also clearly show the contribution of the EPI to the $G$ band linewidth of pristine graphene, and especially it dominates the $\gamma^{\rm in}$ below 500~K. Prior calculations have demonstrated that the EPI in graphene can be tuned over a wide range by changing the carrier density through the application of a gate voltage~\cite{eph1,eph2,Si2013,Park2014,Kim2016,yang2020indirect}. While it is not possible to apply a gate voltage to our suspended monolayer graphene sample, we investigate this effect by calculating the $G$ band linewidth arising from the e-ph scattering for graphene with doping of either electrons or holes.

\begin{figure*}
    \centering
    \includegraphics[width=6.5in]{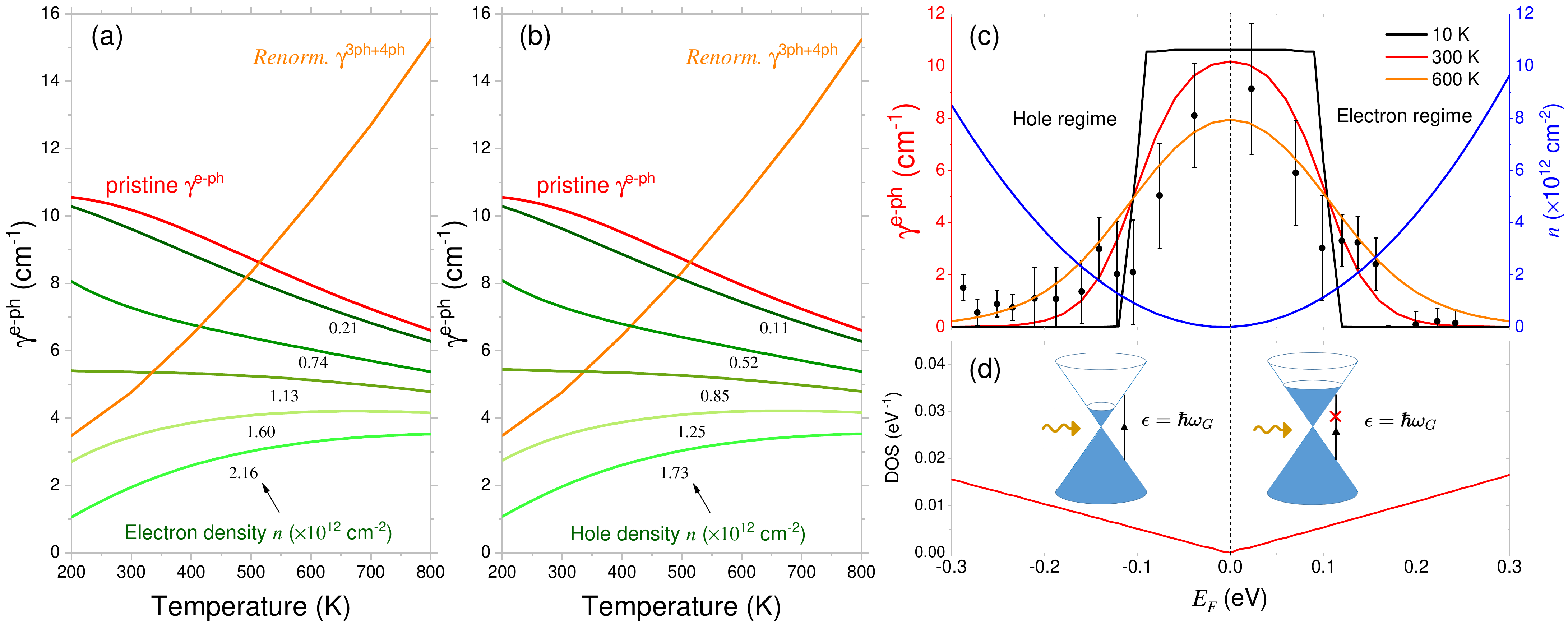}
    \caption{$T$-dependent $G$ band linewidth arising from the EPI $\gamma^{\rm e-ph}$ at different carrier densities for electron-doped graphene (a) and for hole-doped graphene (b). The orange solid lines denote the linewidth contributed by the intrinsic ph-ph scattering ($\gamma^{\rm 3ph}$+$\gamma^{\rm 4ph}$). (c) The $\gamma^{\rm e-ph}$ at different temperatures and carrier density as a function of Fermi energy $E_F$. The vertical dotted line is the position of the charge-neutral Dirac point. The blue solid line is the carrier density; the other solid lines are the calculated $\gamma^{\rm e-ph}$ at three temperatures. The black dots represent experimental data from Ref.~\cite{eph1}. To highlight only the effect of e-ph scattering, we have isolated $\gamma^{\rm e-ph}$ from the original experimental data in Ref.~\cite{eph1}. (d) Electron density of states (DOS) of graphene near the Dirac point. The insets are schematic diagrams for e-ph scattering process applicable to the case of the $G$ mode. $\epsilon$ is the onset energy for vertical electron-hole pair transitions. The left inset represents the decay of the $G$ phonon into electron-hole pairs occurring at low carrier densities ($E_F<\hbar\omega_G/2$ ). The right inset indicates that the $G$ phonon decay into the electron-hole pair is forbidden by the Pauli exclusion principle at high carrier densities ($E_F>\hbar\omega_G/2$).}
    \label{e-ph}
\end{figure*}

Figure~\ref{e-ph}(a, b) shows the calculated $T$-dependent $\gamma^{\rm e-ph}$ of the $G$ band at different carrier densities $n$ for graphene. The contribution from the intrinsic ph-ph scattering ($\gamma^{\rm 3ph}$+$\gamma^{\rm 4ph}$), which is independent of carrier density, is also provided for comparison. It can be seen that $\gamma^{\rm e-ph}$ is significantly decreased with increasing $n$. This indicates that as the carrier density increases, the contribution of e-ph scattering to the linewidth decreases as compared to the contribution from ph-ph scattering. It is also seen in Fig.~\ref{e-ph}(a) that the $T$ dependence of the $\gamma^{\rm e-ph}$ is strongly dependent on the carrier concentration. For low electron densities, e.g., $n=\rm 2.1 \times 10^{11}~cm^{-2}$, the calculated $\gamma^{\rm e-ph}$ decreases with increasing $T$, whereas it increases with $T$ for carrier densities above $\sim \rm 1.1 \times 10^{12} \ cm^{-2}$. From Eq.~\ref{eq.e-p}, the temperature dependence of $\gamma^{\rm e-ph}$ is governed by $f_{\mathbf{k}i}(T)-f_{\mathbf{k}j}(T)$, which are closely associated with the sharpness of the Fermi function and the position of the Fermi energy relative to the threshold of the onset energy for vertical transitions of an electron from a $\pi$ valence band to a $\pi^*$ conduction band state~\cite{eph1,eph2}. This energy corresponds to $\hbar\omega_G\approx0.2$~eV. For low carrier density regimes ($E_F<\hbar\omega_G/2$), as the Fermi function is smeared out with increasing $T$, the number of empty electron states available for transition by absorbing the $G$ phonon is reduced, thus causing the $\gamma^{\rm e-ph}$ to decrease with increasing $T$. As the carrier density increases to reach $E_F>\hbar\omega_G/2$, the smoothing of the Fermi function with increasing $T$ makes part of the occupied electronic states available, and consequently $\gamma^{\rm e-ph}$ increases with $T$. These analyses are also applicable to the case of holes. 

To examine the carrier dependence closely, Fig.~\ref{e-ph}(c) displays $\gamma^{\rm e-ph}$ of the $G$ band varying with $E_F$. The carrier density with respect to the Fermi energy is also included in Fig.~\ref{e-ph}(c) (blue curve). It is remarkable in Fig.~\ref{e-ph}(c) that the variation of $\gamma^{\rm e-ph}$ with doping level varies with $T$. Note that our calculation at 10~K is in reasonable agreement with recent experimental measurements~\cite{eph1}, and that at 300~K coincides with the previous theoretical prediction~\cite{Lazzeri2006}. The difference between experiment~\cite{eph1} and our calculation at 10 K is mainly due to the local density variations in graphene samples. Following the Pauli exclusion principle, near the ground state ($T \rightarrow 0$~K) the vertical transitions can be allowed only when $E_F < 0.1$~eV (corresponding to $\sim \rm 1.1 \times 10^{12} \ cm^{-2}$), as illustrated in the inset of Fig.~\ref{e-ph}(d) on the left, where the energy selection rules are easily satisfied. When $E_F>0.1$~eV, the energy selection rules fully prohibit the $\pi \rightarrow \pi^*$ transitions, as illustrated in the inset of Fig.~\ref{e-ph}(d) on the right. Hence, as seen in Fig.~\ref{e-ph}(c), at $T=10$~K the $\gamma^{\rm e-ph}$ suddenly drops at $E_F=0.1$~eV. As $T$ increases, however, the smoothing of the Fermi function makes part of the occupied electronic states available, thus the $\gamma^{\rm e-ph}$ is smeared out also. As a consequence, the threshold of the Fermi energy, above which $\gamma^{\rm e-ph}$ vanishes, increases with $T$. It can be seen in Fig.~\ref{e-ph}(c) that at $T=300$~K the threshold of the Fermi energy increases to $\sim 0.2$~eV from $0.1$~eV at $T=10$~K, and it reaches $\sim0.3$~eV at $T=600$~K. Note also that as $E_F$ increases, $\gamma^{\rm e-ph}$ exhibits a nearly symmetric reduction relative to the position of the charge-neutral Dirac point, which is closely related to the symmetry of the electron density of states (DOS) near the Dirac point, as shown in Fig.~\ref{e-ph}(d).

\begin{figure}
    \centering
    \includegraphics[width=2.5in]{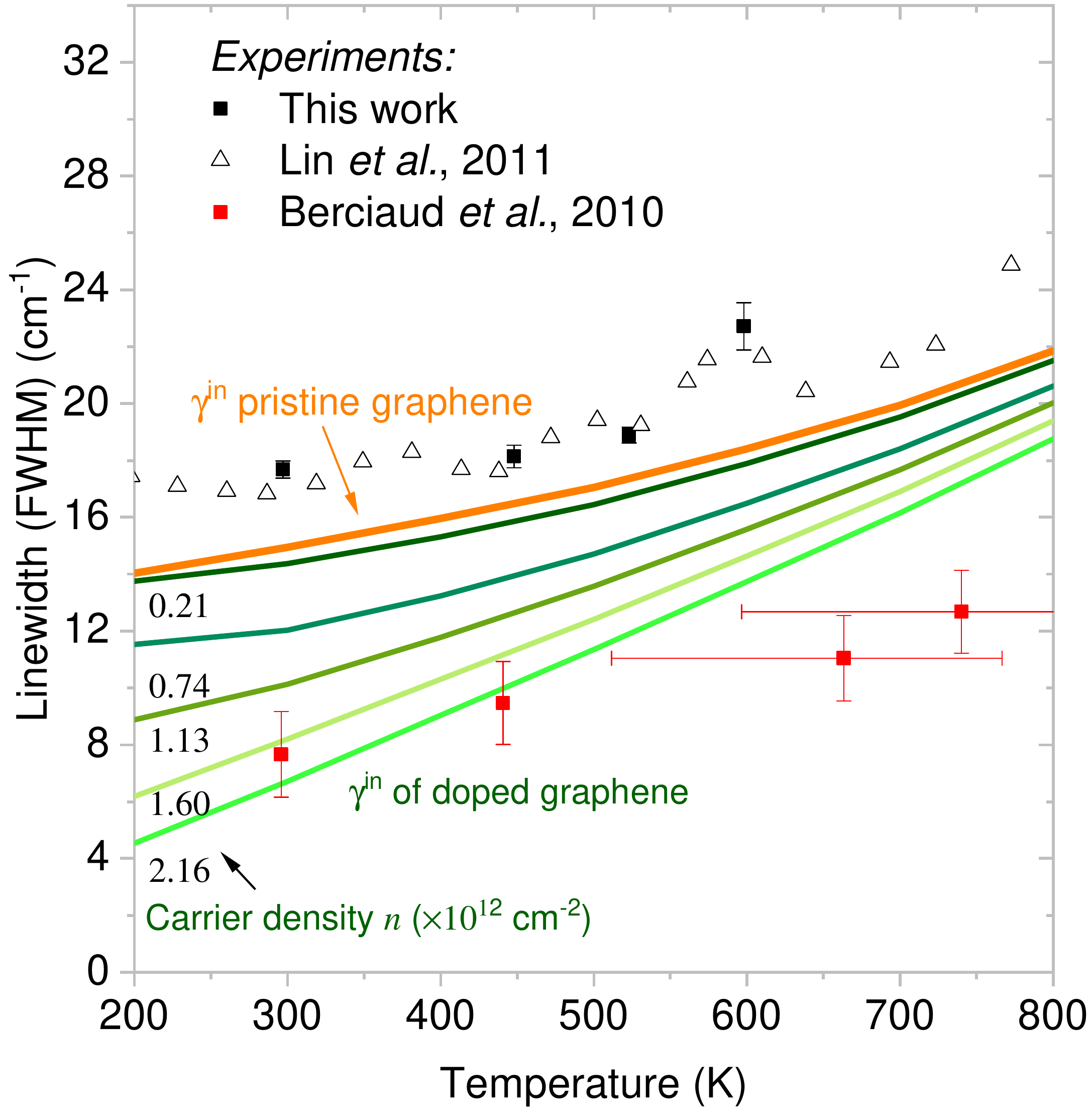}
    \caption{$G$ band linewidth as a function of temperature for different doping levels. The experiment data are from Ref.~\cite{prl2010opExp,prb2011measurement} and our own measurements.}
    \label{gamma-ep}
\end{figure}

Figure~\ref{gamma-ep} shows the calculated total linewidth of the $G$ mode at different carrier concentrations in comparison to available experimental data~\cite{prb2011measurement,prl2010opExp}. Our calculations show that the FWHM of the $G$ mode is extremely sensitive to the carrier density and can vary over a wide range as the carrier density changes. It is clear that after considering a residual charge density of $\rm 2.16 \times 10^{12}~\mathrm{cm^{-2}}$ our calculation can well explain another Raman measurement~\cite{prl2010opExp}. This finding indicates that the variations of the $G$ band linewidth among reported experimental data can be attributed to the e-ph scattering contribution, which strongly depend on the carrier density. On the other hand, the $T$ dependence of the FWHM exhibits a strong doping dependence, which is strongly connected to the interplay between the ph-ph scattering and e-ph scattering. At low carrier densities, e.g., $n=\rm 2.1 \times 10^{11}~\mathrm{cm^{-2}}$, the calculated linewith increases slowly with $T$, since e-ph processes make a dominant contribution to $\gamma$ and it partially compensates the increase of the linewidth due to ph-ph processes. At higher carrier densities,  e.g., $n=\rm 2.16 \times 10^{12}~\mathrm{cm^{-2}}$, the FWHM is completely dominated by the ph-ph processes, and it thus increase rapidly as $T$ increases. These results reveal the significance of the e-ph scattering in determining the $G$ band linewith and its temperature dependence in graphene.

In conclusion, we have investigated the $G$ band frequency shift and linewith of graphene by including phonon renormalization and the anharmonic three-phonon, four-phonon, and electron-phonon scattering contributions, using first principles. We reveal that four-phonon scattering, which was neglected in the past, plays an indispensable role in the $G$ band linewith of pristine graphene, although it is greatly weakened by phonon renormalization. When combining both ph-ph and e-ph scattering, our prediction successfully explains previous measurements and our own measurements. Our calculation also shows that the e-ph coupling contribution and its temperature dependence significantly varies with doping levels. By calculating the $G$ band linewidth at different carrier densities, we suggest that the variations among previous experiments results can be attributed to the e-ph coupling contribution. Our work provides important insights into the understanding phonon-phonon interaction and electron-phonon interaction of graphene and other 2D materials. \\


\begin{acknowledgments}
X. R., Z. H., and L. S. were supported by two collaborating grants (2015946 and 2015954) of the US National Science Foundation. Simulations were performed at the Rosen Center for Advanced Computing (RCAC) of Purdue University. X.Y. was supported by the Natural Science Foundation of China (Grant No. 12004254). W. L. was supported by the GuangDong Basic and Applied Basic Research Foundation (Grant No. 2021A1515010042) and the Shenzhen Science, Technology and Innovation Commission (Grant No. 20200809161605001). We would like to thank Professor Xiaokun Gu for the helpful discussions on thermal expansion and temperature-dependent force constants. 
\end{acknowledgments}



\bibliography{Reference}

\end{document}


\title{Supplemental Material for ``Raman Linewidth Contributions from Four-Phonon and Electron-Phonon Interactions in Graphene"}

\author{Zherui Han}
\altaffiliation{These authors contributed equally to this work.}
\affiliation{School of Mechanical Engineering and the Birck Nanotechnology Center,\\
	Purdue University, West Lafayette, Indiana 47907-2088, USA.}
\author{Xiaolong Yang}%
\altaffiliation{These authors contributed equally to this work.}
\affiliation{Institute for Advanced Study, Shenzhen University, Shenzhen 518060, China.}
\author{Sean Sullivan}
\affiliation{Department of Mechanical Engineering, The University of Texas at Austin, Austin, TX 78712, USA.}
\author{Tianli Feng}
\affiliation{Department of Mechanical Engineering, University of Utah, Salt Lake City, Utah 84112, USA.}
\author{Li Shi}
\affiliation{Department of Mechanical Engineering, The University of Texas at Austin, Austin, TX 78712, USA.}
\author{Wu Li}
\email{wu.li.phys2011@gmail.com}
\affiliation{Institute for Advanced Study, Shenzhen University, Shenzhen 518060, China.}
\author{Xiulin Ruan}%
\email{ruan@purdue.edu}
\affiliation{School of Mechanical Engineering and the Birck Nanotechnology Center,\\
	Purdue University, West Lafayette, Indiana 47907-2088, USA.}

\date{\today}

\maketitle
In this supplemental material, we cover the compurational details of our study.

All calculations are done using Density Functional Theory (DFT) or Density Functional Perturbation Theory (DFPT). For phonon-phonon interactions in pristine graphene, we employ VASP package~\cite{VASP1993} and use Perdew-Burke-Ernzerhof (PBE) parameterization of the generalized gradient approximation (GGA) for exchange and correlation functionals~\cite{prl1996GGA}. The plane wave cutoff is 600~eV. We also construct $8 \times 8 \times 1$ supercells and use $3 \times 3 \times 1$ $k$-mesh to calculate interatomic force constants (IFCs) and consider the tenth and second nearest neighboring atoms for third-order IFCs and fourth-order IFCs, respectively. Brillouin Zone (BZ) is discretized by $150 \times 150 \times 1$ $q$-mesh to evaluate phonon scattering rates using ShengBTE package~\cite{shengbte} for three-phonon and $50 \times 50 \times 1$ $q$-mesh using in-house codes (which is recently made public~\cite{han2021fourphonon}) for four-phonon. 

To consider the phonon renormalization effect in graphene, we use TDEP to get temperature-dependent (or effective) IFCs. This arises from the consideration that harmonic approximation may fall short when describing highly anharmonic systems and the vibrational properties at finite-temperatures are not captured correctly. In recent years, thermal science community has developed and refined methods to treat this temperature effect, which now are able to study both harmonic and anharmonic properties at certain temperatures~\cite{TDEP2013,xia2018revisiting,ravichandran2018unified}. In this work, we choose TDEP developed by Hellman et al., since it is compatible with ab initio calculations and can simulate thermal displacements of atoms using Bose–Einstein distribution~\cite{TDEP2013,TDEP2013IFCs}. The key of this method is to obtain effective IFCs at finite-temperatures that can fit into forces-displacements data. We employ a $10 \times 10 \times 1$ supercell of graphene and iterate the calculations using 100 thermally perturbed snapshots. At each temperature, the last iteration is done using 400 snapshots to ensure convergence and three iterations are sufficient in our calculations.

For electron-phonon interactions, all calculations are done using QUANTUM-ESPRESSO package~\cite{QE} with local-density approximation, norm-conserving pseudopotentials~\cite{nc-potential}, and a kinetic energy cutoff of 60~Ry. The EPW package~\cite{EPW} is employed to perform Wannier function interpolation for the e-ph coupling matrix elements. The related quantities are calculated first on a coarse grid of $8 \times 8 \times 1$ and then Wannier interpolated into a fine gird of $700\times700\times1$ for $\gamma^{\rm e-ph}$. The electronic integration over the BZ is approximated by the Gaussian smearing of 0.025 Ry for the self-consistent calculations. Doped graphene is modeled by shifting the Fermi surface $E_F$ towards the unoccupied energy states from the charge neutral point.

\bibliography{supplement}